\documentclass[aps,prl,twocolumn,groupedaddress]{revtex4}
\usepackage[dvips]{graphicx,color}
\usepackage{amsmath,amssymb}

\begin{document}

\newcommand{\etal}      {{\it et~al.}}


\title{Direct Observation of the Hyperfine Transition of the Ground State Positronium}



\author{T. Yamazaki,$^1$ A. Miyazaki,$^1$ T. Suehara,$^1$ T. Namba,$^1$ S. Asai,$^1$ T. Kobayashi,$^1$ H. Saito,$^2$ I. Ogawa,$^3$ T. Idehara,$^3$ and S. Sabchevski$^4$}
\affiliation{$^1$Department of Physics, Graduate School of Science, and International Center for Elementary Particle Physics, University of Tokyo, 7-3-1 Hongo, Bunkyo-ku, Tokyo 113-0033, Japan}
\affiliation{$^2$Institute of Physics, Graduate School of Arts and Sciences, University of Tokyo, 3-8-1 Komaba, Meguro-ku, Tokyo 153-8902, Japan}
\affiliation{$^3$Research Center for Development of Far-Infrared Region, University of Fukui, 3-9-1 Bunkyo, Fukui-shi, Fukui 910-8507, Japan}
\affiliation{$^4$Bulgarian Academy of Sciences, 72 Tzarigradsko Shose Blvd., 1784 Sofia, Bulgaria}


\date{\today}

\begin{abstract}
We report the first direct measurement of the hyperfine transition of the ground state positronium.
The hyperfine structure between ortho-positronium  and para-positronium is about 203~GHz.
We develop a new optical system to accumulate about 10~kW power using a gyrotron, a mode converter, and a Fabry-P\'{e}rot cavity.
The hyperfine transition has been observed with a significance of 5.4 standard deviations.
The transition probability is measured to be $A = 3.1^{+1.6}_{-1.2} \times 10^{-8}$~s$^{-1}$ for the first time, which is in good agreement with the theoretical value of $3.37 \times 10^{-8}$~s$^{-1}$.
\end{abstract}


\maketitle


Positronium (Ps)~\cite{Ps-REV}, a bound state of an electron and a positron, is a purely leptonic system and is a good target to study quantum electrodynamics (QED) in bound state.
The triplet ($1^{3}S_{1}$) state of Ps, ortho-positronium (o-Ps), decays into three gamma rays with a lifetime of $\tau_{\text{o}} = 142$~ns~\cite{LIFE-TOKYO, LIFE-MICHIGAN}.
On the other hand, the singlet ($1^{1}S_{0}$) state of Ps, para-positronium (p-Ps), decays into two gamma rays in $\tau_{\text{p}} = 125$~ps~\cite{LIFE-PARA}.
The energy level of the ground state o-Ps is higher than that of the ground state p-Ps due to the spin-spin interaction between the electron and the positron.
This difference is called the hyperfine structure of the ground state positronium (Ps-HFS), which is about 203~GHz.
Although precise measurements of Ps-HFS have been performed in 1970s and 1980s~\cite{HFS-MILLS,HFS-RITTER}, all of them are indirect measurements using Zeeman splitting of about 3~GHz caused by a static magnetic field of about 1~T.
There is a discrepancy of 3.9 standard deviations (15~ppm) between the measured and the theoretical value~\cite{HFS-THEORY}.
The largest systematic uncertainty common to all previous measurements is the non-uniformity of the static magnetic field.
It is important to directly measure Ps-HFS, in order to avoid the systematic uncertainty of the static magnetic field.
Here we present a direct observation of the hyperfine transition between Ps-HFS, which is the first great step toward a direct measurement of Ps-HFS.
The hyperfine transition of the ground state Ps, which is $M1$ transition, has not yet been observed directly, since the transition probability (Einstein's $A$ coefficient is $A = 3.37 \times 10^{-8}$~s$^{-1}$~\cite{A-THEORY})  is $10^{14}$ times smaller than the decay rate of o-Ps ($7.040 1(6) \times 10^{6}$~s$^{-1}$~\cite{LIFE-TOKYO, LIFE-MICHIGAN}).
In order to cause sufficient amount of stimulated emission from o-Ps to p-Ps, we develop a new optical system which consists of a gyrotron as a sub-THz radiation source, a mode converter to convert the gyrotron output to a Gaussian beam, and a Fabry-P\'{e}rot cavity to accumulate high power sub-THz radiation.
The gyrotron is a novel high power radiation source for sub-THz to THz region, which enables us to perform a direct measurement of the hyperfine transition.
High power 203~GHz radiation in the Fabry-P\'{e}rot cavity causes the hyperfine transition from the ground state o-Ps to p-Ps, and p-Ps promptly decays into two back-to-back 511~keV gamma rays.
Consequently, the transition signal ($\text{o-Ps}\rightarrow\text{p-Ps}\rightarrow2\gamma$) has distinctive features that it has a lifetime of o-Ps and decays into two back-to-back 511~keV gamma rays as p-Ps.

\begin{figure}
\includegraphics[width=85mm]{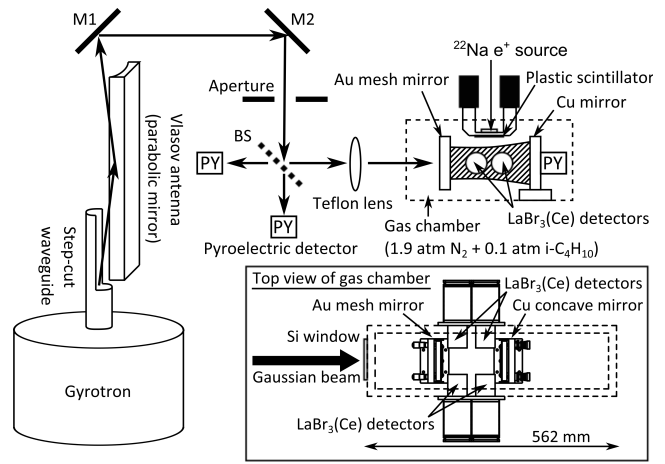}
\caption{Schematic diagrams of our experimental setup. 
Top view of the gas chamber is shown in the box.
M1 and M2 are parabolic mirrors made of aluminum.
We use a gold mesh plane mirror with a transmittance of about 3~\% as a beam splitter (BS).
Three pyroelectric detectors (PY) are used to monitor the incident, the reflected and the transmitted power.
\label{fig:setup}}
\end{figure}

Figure \ref{fig:setup} shows a schematic view of our experimental setup.
We use Gyrotron FU CW V~\cite{FU-CW-V}, which produces 202.89~GHz (140.06~GHz) radiation in TE$_{03}$ (TE$_{02}$) mode in 15~ms pulses at 20~Hz.
The power is monitored with a pyroelectric detector, which is fed back to voltage of the heater of the electron gun.
As a result, it can operate stably with about 300~W power within 10~\% fluctuation.

In order to enhance the output power of the gyrotron, the radiation is accumulated in a Fabry-P\'{e}rot cavity.
The gyrotron output (TE$_{0n}$ mode) is converted to a Gaussian beam so as to obtain good coupling with the Fabry-P\'{e}rot cavity.
Main components of a mode converter are a step-cut waveguide and a large parabolic mirror made of aluminum (Vlasov antenna).
They convert TE$_{0n}$ mode to a bi-Gaussian beam geometrically if the axis of the step-cut waveguide and the focal point of the parabola are matched~\cite{VLASOV-OGAWA}.
Two mirrors (M1 and M2) are used to convert the bi-Gaussian beam into a Gaussian beam.
In order to improve the beam quality, we insert an aperture (diam.~$= 50$~mm) as a spatial filter to block out side lobes of the beam.
Spatial distribution of the beam is measured by exposing a PVC sheet to the beam and taking its picture by an infrared camera.
Power conversion efficiency, which is estimated from the spatial distribution, is $28\pm 2$~\%  due to a limitation of a purity of wave mode in the gyrotron output.

The Fabry-P\'{e}rot cavity is made with a gold mesh plane mirror (diam.~$= 50$~mm) and a copper concave mirror (diam.~$= 50$~mm, curvature $= 300$~mm).
The incident Gaussian beam resonates within the Fabry-P\'{e}rot cavity when the cavity length (136~mm) is equal to a half-integer multiple of the wavelength of the radiation (about 1.5~mm).
The cavity length is controlled by moving the copper concave mirror mounted on an X-axis stage (NANO CONTROL TS102-G).
The gold mesh plane mirror is a key component of the Fabry-P\'{e}rot cavity, and is made on a SiO$_{2}$ plate using photolithography and liftoff technique.
The line width and separation are 200~$\mu$m and 160~$\mu$m, respectively.
The mesh parameters are designed to obtain high reflectivity (99.38~\%) and reasonable transmittance (0.39~\%), which are simulated with CST Microwave Studio~\cite{CST}.
As a result, the finesse of the Fabry-P\'{e}rot cavity attains ${\mathcal F} = 623 \pm 29$, which is estimated from the width of the resonance peak while changing cavity length. 
The power accumulated in the Fabry-P\'{e}rot cavity reaches about 10~kW.

The power accumulated in the Fabry-P\'{e}rot cavity is estimated with the power transmitted through a hole (diam.~$= 0.6$~mm) at the center of the copper concave mirror.
The transmitted power is monitored with a pyroelectric detector.
A ratio between accumulated power and transmitted power is obtained from independent measurements using the Gaussian beam as follows.
First, the beam undergoes total absorption in water.
Its total power is estimated from a temperature increase of the water.
Next, the copper concave mirror is exposed to the beam, and power transmitted through the hole is measured with the pyroelectric detector.
From these measurements, the ratio of the Gaussian beam power to transmitted power is obtained.
However, the spatial distribution of the Gaussian beam is different from that of the beam inside the Fabry-P\'{e}rot cavity.
Correcting the difference of the beam shapes and considering that only the beam going to the copper mirror direction can be transmitted through the hole, the ratio of accumulated power to transmitted power is obtained.
The Gaussian beam shape is measured with a PVC sheet, and the spatial distribution in the Fabry-P\'{e}rot cavity is calculated from the cavity length, the curvature of the copper mirror, and the radiation wavelength.
The uncertainty of the ratio of accumulated power to transmitted power is $^{+33}_{-30}$~\% because of the fluctuation of the beam shape between the exposure on the copper mirror and that on the PVC sheet.
The uncertainty of power does not affect the direct measurement of the hyperfine transition, but contributes to the accuracy of the transition probability, which is also measured in our experiment.

\begin{figure}
\includegraphics[width=85mm]{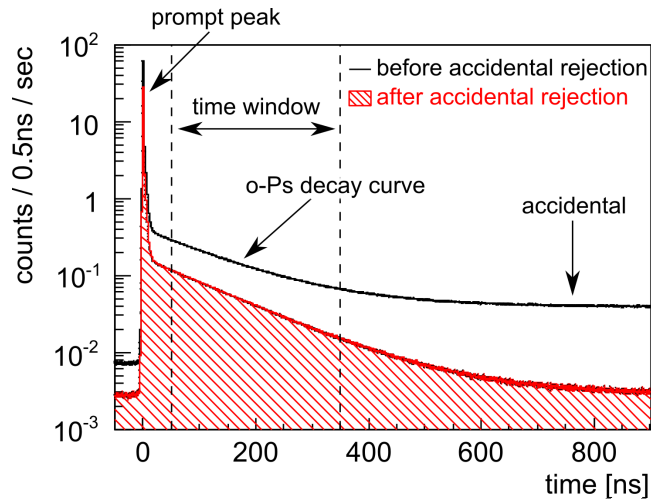}
\caption{Time difference between the plastic scintillator and the coincidence signal of the LaBr$_3$(Ce) scintillators.
Solid line and hatched histogram show the time spectrum before and after accidental rejection cut, respectively. 
The time window for delayed coincidence is shown as a dashed line.\label{fig:tspec}}
\end{figure}

Positronium formation assembly shown in Fig.~\ref{fig:setup} is as follows:
A 780 kBq $^{22}$Na positron source is placed above a thin plastic scintillator (NE-102, thickness $= 0.1$~mm). 
Emitted positrons pass through the scintillator and produce light pulses that are directed to two 1.5-inch fine-mesh photomultipliers (Hamamatsu R5924-70) by the light guide.
Positrons form Ps when stopped in the mixed gas (1.9~atm N$_{2}$ and 0.1~atm i-C$_{4}$H$_{10}$)~\cite{Ps-FORM}.
About 5~\% of positrons are tagged by the plastic scintillator and stop in the gas, and then, about 1/4 of them form Ps.
Therefore, the Ps formation rate is about 10$^{4}$~s$^{-1}$.
Ps has kinematic energy of about 1~eV just after its formation.
It becomes thermalized after $O(10\text{~ns})$ with elastic collisions with gas molecules and the kinetic energy becomes about 1/30~eV.
Since we use delayed coincidence as shown in Fig.~\ref{fig:tspec}, the width of the Doppler broadening due to motion of thermalized Ps is only about $\Delta f_{\text{D}} = 0.08$~GHz, which is much smaller than the natural linewidth $\Delta f_{\text{n}} \sim 1/2\pi\tau_{\text{p}} = 1.27$~GHz.

Gamma rays emitted from Ps decay are observed in four LaBr$_{3}$(Ce) crystals (Saint-Gobain Crystals, diam.~$= 1.5$~inch and length = 2.0~inch).
The four detectors are placed as shown in Fig.~\ref{fig:setup} to make four back-to-back pairs.
The scintillation pulses of the LaBr$_3$(Ce) crystals are detected with 1.5-inch fine-mesh photomultipliers (Hamamatsu R5924-70).
The energy resolution of the LaBr$_{3}$(Ce) detectors is 4~\% (FWHM) at 511~keV.
The primary decay time is 16~ns.
These are advantages for tagging monochromatic 511~keV gamma rays and avoiding pileup of gamma ray signals.

In order to select Ps decay events, data acquisition logic is set up
as follows:
When at least one back-to-back signal from the LaBr$_3$(Ce) scintillator pairs is coincident within 40~ns, and then when this coincidence is within $-$100~ns to 1100~ns of the timing of the plastic scintillator, data acquisition is triggered.
A charge ADC (PHILLIPS 7167) and another charge ADC (REPIC RPC-022) are used to measure the energy information of the plastic scintillator with short and long gate, respectively.
The energy difference between short and long gates is used to suppress accidental background as mentioned later.
The outputs of the LaBr$_3$(Ce) detectors are recorded with a charge ADC (CAEN C1205).
The time information between the plastic and LaBr$_3$(Ce) scintillators is recorded using a direct clock (2~GHz) count type TDC (KEK GNC-060)~\cite{LIFE-TOKYO}.

Four runs have been performed.
In three runs (Run I, III, and IV), 202.89~GHz radiation (TE$_{03}$ mode) is used, and different powers are accumulated in the Fabry-P\'{e}rot cavity (11.0~kW, 0.0~kW, and 5.6~kW).
In another run (Run II), off-resonance frequency of 140.06~GHz in TE$_{02}$ mode is used to check systematic effects due to the absorption of the radiation in the mixed gas.
Total period of data acquisition is about two weeks.
During the data acquisition, energy and time calibrations are performed every 30 minutes.
Trigger rates are about 1~kHz.
The $\gamma$-ray peak at 511~keV and the zero energy peak are used to calibrate the LaBr$_3$(Ce) detectors. 
The room temperature is maintained within $26 \pm 1\ ^{\circ}$C in order to maintain good stability during the data acquisition.

Figure \ref{fig:tspec} shows the time difference between the plastic scintillator signal and the coincidence signal of the LaBr$_3$(Ce) detectors.
A sharp peak from prompt annihilation is followed by the exponential curve of transition signals and o-Ps decay signals, and then the constant spectrum due to accidental overlaps of a triggered positron and uncorrelated gamma rays.
A good timing resolution ($\sigma = 0.8$~ns) is obtained.
After selecting a time window from 50~ns to 350~ns to enhance the transition signals and o-Ps decay events, accidental events remain as the dominant source of back-to-back 511~keV gamma rays. 
In the case of accidental events, there is another plastic scintillator hit at the timing of $\gamma$-ray hit.
The energy deposit on the plastic scintillator measured with long gate becomes larger than that measured with short gate.
To reject accidental events, the energy difference between long gate and short gate is limited from $-2.5$~pe (photoelectron) to 1.7~pe.
This cut is applied on both PMTs of the plastic scintillators.

\begin{figure}
\includegraphics[width=85mm]{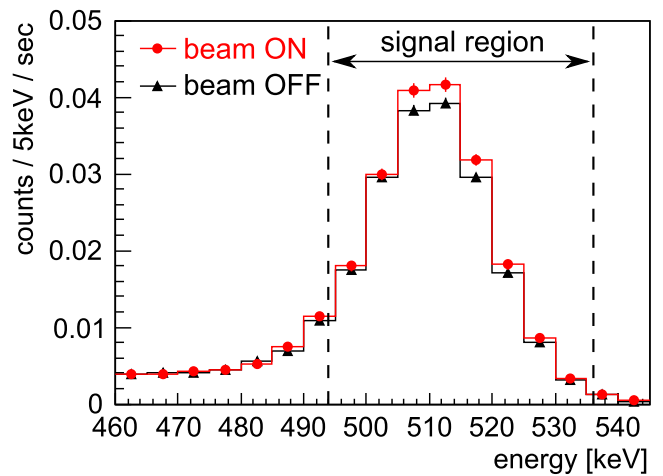}
\caption{Energy spectra of the LaBr$_3$(Ce) scintillator in the highest power on-resonance run (Run I, $11.0^{+3.6}_{-3.3}$~kW) after all event selections are applied.
Circles and triangles show ``beam ON'' and ``beam OFF'' spectra, respectively.\label{fig:espec}}
\end{figure}

Finally, we count the number of events in which back-to-back 511~keV gamma rays are observed.
Figure~\ref{fig:espec} shows the energy spectra measured with the LaBr$_3$(Ce) scintillator in the highest power on-resonance run (Run I, $11.0^{+3.6}_{-3.3}$~kW).
The delayed coincidence and the accidental rejection are applied.
In addition, a 511~keV $\gamma$-ray hit on the LaBr$_3$(Ce) scintillator at the opposite side of the back-to-back pair is required, where the energy window is set from 494~keV to 536~keV.
Remaining accidental background is estimated from the events in another time window set from 850~ns to 900~ns, and is subtracted.
Circles and triangles show ``beam ON'' and ``beam OFF'' spectra, respectively.
The data taken during ``beam OFF'' period in the pulse beam are used to estimate background.
The ``beam OFF'' spectrum consists of pick-off annihilation ($\text{o-Ps}+e^{-}\rightarrow2\gamma+e^{-}$) and $3\gamma$ decay ($\text{o-Ps}\rightarrow3\gamma$) of o-Ps.
Transition signals ($\text{o-Ps}\rightarrow\text{p-Ps}\rightarrow2\gamma$) increase when o-Ps are exposed to high power sub-THz radiation during ``beam ON'' period.
The signal rate in the energy window from 494~keV to 536~keV is $R_{\text{ON}}-R_{\text{OFF}} = 15.1 \pm 2.7 ({\text{stat.}})$~mHz, where $R_{\text{ON}}$ ($R_{\text{OFF}}$) is the ``beam ON'' (``beam OFF'') event rate after all event selections are applied.

Systematic errors are summarized in Table~\ref{tab:sys}.
The largest contribution is the uncertainty in Ps formation probability.
Ps formation probabilities of the ``beam ON'' and the ``beam OFF'' data are different because of absorption of the sub-THz radiation in the mixed gas, which is enhanced when the beam resonates with the Fabry-P\'{e}rot cavity.
The difference is estimated by counting the number of events in the time window before the energy cut is applied, since Ps formation probability is independent of the $\gamma$-ray energy cut.
The difference in Ps formation probability is the largest in off-resonance run (Run II).
Another dominant systematic error is uncertainty in the efficiency of the accidental rejection cut.
Inefficiency of the accidental rejection depends on the rates of the plastic scintillator signals which go over the discriminator threshold ($\sim 1$~pe).
This systematic effect is estimated from the difference of the efficiency of the accidental rejection between ``beam ON'' and ``beam OFF'', which is independent of the $\gamma$-ray energy cut.
In addition, if the energy resolution and energy scale of the LaBr$_3$(Ce) scintillator are different between ``beam ON'' and ``beam OFF'', fake signals appear because of the back-to-back 511~keV energy selection.
This effect is estimated using Monte Carlo simulation with Geant4~\cite{GEANT4} where the energy resolution and energy scale taken from data are used as input.
The last dominant source is the uncertainty of background normalization.
The background is estimated from ``beam OFF'' events.
Its normalization is performed using the number of events in the prompt time window set from $-3$~ns to 1.5~ns, where the usual $e^{+}$ annihilation is dominant (77~\%).
Statistical accuracy determines the normalization uncertainty.

\begin{table}
\caption{Summary of the systematic errors. The values are ratios to the background.\label{tab:sys}}
\begin{ruledtabular}
\small
\begin{tabular}{lcccc}
\multicolumn{1}{c}{Source} & Run I        & Run II       & Run III      & Run IV       \\ \hline
Ps formation prob.         & $-0.27$~\%   & $-0.39$~\%   & $+0.20$~\%   & $-0.13$~\%   \\
Accidental rejection       & $+0.17$~\%   & $+0.05$~\%   & $+0.13$~\%   & $+0.23$~\%   \\
Energy resolution          & $-0.08$~\%   & $+0.06$~\%   & $-0.11$~\%   & $-0.02$~\%   \\
BG normalization           & $\pm0.03$~\% & $\pm0.04$~\% & $\pm0.04$~\% & $\pm0.04$~\% \\
Total                      & $^{+0.17}_{-0.29}$~\% & $^{+0.08}_{-0.39}$~\% & $^{+0.24}_{-0.12}$~\% & $^{+0.24}_{-0.14}$~\%
\end{tabular}
\end{ruledtabular}
\end{table}

\begin{figure}
\includegraphics[width=80mm]{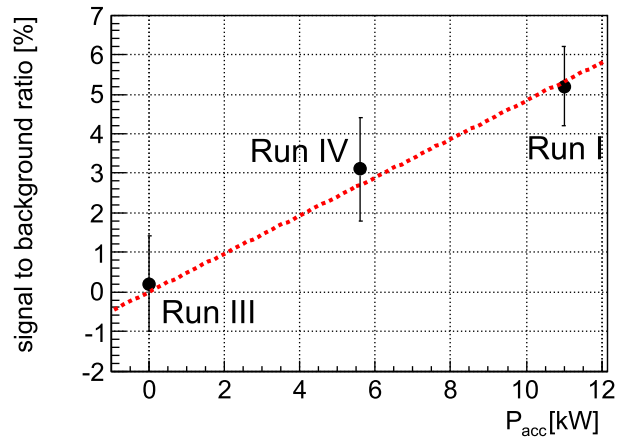}
\caption{Power dependence of the amount of the transition signals. The vertical axis shows signal to background ratio. The horizontal axis shows the power accumulated in the Fabry-P\'{e}rot cavity. The error bars contain statistical uncertainty as well as systematic uncertainties summarized in Table~\ref{tab:sys}. The dashed line shows the result of a linear fit.\label{fig:sn_power}}
\end{figure}

The systematic errors discussed above are independent, and the total systematic error can be calculated as their quadrature sum. 
Final result with the systematic errors is 
\begin{equation}\label{eq:sigrate}
R_{\text{ON}} - R_{\text{OFF}} = 15.1 \pm 2.7 ({\text{stat.}}) ^{+0.5}_{-0.8} ({\text{sys.}})\ \text{mHz}.
\end{equation}
This is the first direct observation of the hyperfine transition of the ground state positronium with a significance of 5.4 standard deviations.
In addition, the fraction of the transition signals is proportional to the power accumulated in the Fabry-P\'{e}rot cavity (Fig.~\ref{fig:sn_power}), and the off-resonance data (Run II) gives a null result as expected, despite the relatively large difference in Ps formation probability as seen in Table~\ref{tab:sys}.

The transition probability (or Einstein's $A$ coefficient) between the ground state Ps-HFS is also measured for the first time.
It can be estimated from the observed transition rate, the power accumulated in the Fabry-P\'{e}rot cavity, and $2\gamma / 3\gamma$ detection efficiency estimated from Monte Carlo Simulation with Geant4.
The estimated result is
\begin{equation}\label{eq:A}
A = 3.1^{+1.6}_{-1.2} \times 10^{-8}\ \text{s}^{-1},
\end{equation}
which is consistent with the theoretical value of $3.37 \times 10^{-8}$~s$^{-1}$~\cite{A-THEORY}.
The largest uncertainty is the estimation of the absolute power accumulated in the Fabry-P\'{e}rot cavity.

Our next target is to directly measure Ps-HFS for the first time.
Output frequency of gyrotron can be changed with cavities of different sizes.
In Ps-HFS measurement, relative accuracy of the power estimation at different frequency points is necessary.
In addition, in order to perform precise measurement of Ps-HFS, we need more statistics.
A possible way to increase statistics is to use a slow positron beam and make positroniums in vacuum using a thin metal foil~\cite{FOIL}.
It also eliminates systematic uncertainty and beam power loss due to absorption of the sub-THz radiation.

In summary, the hyperfine transition of the ground state positronium has been observed directly for the first time with a significance of 5.4 standard deviations.
We develop a new optical system to accumulate about 10~kW power using a gyrotron, a mode converter, and a Fabry-P\'{e}rot cavity, in order to cause observable amount of stimulated emission from o-Ps to p-Ps.
The transition probability (or Einstein's $A$ coefficient) is also measured to be $A = 3.1^{+1.6}_{-1.2} \times 10^{-8}$~s$^{-1}$ for the first time, which is in good agreement with the theoretical value.

This research was funded in part by the Japan Society for the Promotion of Science. 



\begin{thebibliography}{100}

\bibitem{Ps-REV}
A. Rich, Rev.~Mod.~Phys.~{\bf 53}, 127 (1981). 

\bibitem{LIFE-TOKYO}
S. Asai, S. Orito, and N. Shinohara, Phys.~Lett.~B {\bf 357}, 475 (1995); O. Jinnouchi, S. Asai, and T. Kobayashi, {\it ibid.}~{\bf 572}, 117 (2003); Y. Kataoka, S. Asai, and T. Kobayashi, {\it ibid.}~{\bf 671}, 219 (2009). 

\bibitem{LIFE-MICHIGAN}
R. S. Vallery, P. W. Zitzewitz, and D. W. Gidley, Phys.~Rev.~Lett.~{\bf 90}, 203402 (2003). 

\bibitem{LIFE-PARA}
A. H. Al-Ramadhan and D. W. Gidley, Phys.~Rev.~Lett.~{\bf 72}, 1632 (1994). 

\bibitem{HFS-MILLS}
A. P. Mills, Phys.~Rev.~A {\bf 27}, 262 (1983). 

\bibitem{HFS-RITTER}
M. W. Ritter, P. O. Egan, V. W. Hughes, and K. A. Woodle, Phys.~Rev.~A {\bf 30}, 1331 (1984). 

\bibitem{HFS-THEORY}
B. A. Kniehl, and A. A. Penin, Phys.~Rev.~Lett.~{\bf 85}, 5094 (2000). 

\bibitem{A-THEORY}
P. Wallyn, W. A. Mahoney, P. H. Durouchoux, and C. Chapuis, Astrophys.~J. {\bf 465}, 473 (1996).

\bibitem{FU-CW-V}
T. Idehara and S. P. Sabchevski, J. Infrared Milli.~Terahz.~Waves, online first, doi: 10.1007/s10762-011-9862-x. 

\bibitem{VLASOV-OGAWA}
I. Ogawa \etal, Int.~J. Elec.~{\bf 83}, 635 (1997). 

\bibitem{CST}
CST Microwave Studio 2011, CST Computer Simulation Technology AG, http://www.cst.com

\bibitem{Ps-FORM}
M. Charlton, Rep.~Prog.~Phys.~{\bf 48}, 737 (1985). 

\bibitem{GEANT4}
A. Agostinelli \etal, Nucl.~Instrum.~Meth.~A {\bf 506}, 250 (2003). 

\bibitem{FOIL}
P. J. Schultz and K. G. Lynn, Rev.~Mod.~Phys.~{\bf 60}, 701 (1988).

\end{thebibliography}
\end{document}